\documentclass[twocolumn,showpacs,preprintnumbers,amsmath,amssymb]{revtex4}

\usepackage{graphicx}
\usepackage{dcolumn}
\usepackage{bm}
\usepackage{latexsym}
\usepackage{amsfonts}
\usepackage{amssymb}
\usepackage{amsmath}

\begin{document}


\title{Kaluza-Klein two brane worlds cosmology at low energy}
\author{S.~Feranie$^{(1)}$}
\email{feranie@upi.edu}
\author{Arianto$^{(2,4)}$}
\email{arianto@upi.edu}
\author{Freddy P.~Zen$^{(3,4)}$}
\email{fpzen@fi.itb.ac.id}
\affiliation{$^{(1)}$Jurusan Pendidikan Fisika, FPMIPA,\\
Universitas Pendidikan Indonesia\\
Jl. Dr. Setiabudi 229, Bandung 40154, Indonesia.\\
$^{(2)}$Department of Physics, Faculty of Mathematics and Natural Sciences,\\ Udayana University\\
Jl. Kampus Bukit Jimbaran Kuta-Bali 80361, Indonesia.\\
$^{(3)}$Theoretical Physics Lab., THEPI Division, and \\
$^{(4)}$Indonesia Center for Theoretical and Mathematical Physics (ICTMP)\\
Faculty of Mathematics and Natural Sciences,\\
 Institut Teknologi Bandung,\\
Jl. Ganesha 10 Bandung 40132, Indonesia. }


\begin{abstract}
We study two $(4+n)$-dimensional branes embedded in
$(5+n)$-dimensional spacetime.  Using the gradient expansion
approximation, we find that the effective theory is described by
the $(4+n)$-dimensional scalar-tensor gravity with a specific
coupling function. Based on this theory we investigate the
Kaluza-Klein two brane worlds cosmology at low energy. We study in
both the static and the non-static internal dimensions. In the
static case the effective gravitational constant in the induced
Friedmann equation depends on the equations of state of the brane
matters and the dark radiation term naturally appear. In the
non-static case we take a relation between the external and
internal scale factors of the form $b(t)=a^\gamma(t)$ in which the
brane world evolves with two scale factors. In this case, the
induced Friedmann equation on the brane is modified in the
effective gravitational constant and the term proportional to
$a^{-4\beta}$. For dark radiation, we find $\gamma=-2/(1+n)$.
Finally, we discuss the issue of conformal frames which naturally
arises with scalar-tensor theories. We find that the static
internal dimensions in the Jordan frame may become non-static in
the Einstein frame.
\end{abstract}

\maketitle

\section{Introduction}
One of the most interesting and surprising aspects of the string
theory or M-theory is the fact that it can only be correctly
formulated in a higher dimensional spacetime. On the other hand,
our observed Universe is a four-dimensional spacetime. Therefore
we need a mechanism of compactification of the extra dimensions,
so that they become invisible at least at low energy scales.
Moreover, investigations of non-perturbative string theory has
lead to the discovery that string theory must contain higher
dimensional extended objects called branes. The existence of these
branes has inspired a new method of compactification of extra
dimensions, so that they become invisible at least at low energy
scales. Previously the preferred method was Kaluza-Klein
compactification, in which the extra dimensions are compact and
extremely small. This method of compactification has further
inspired a class of classical models of the universe, in which
extra dimensions can be included in general relativity, and their
possible implications for classical cosmology can be investigated
phenomenologically without any dependence on a particular model of
string theory. This is known as the brane world scenario, in which
the standard particles or fields are confined to a brane, while
the graviton can propagates into the bulk as well as into the
brane. Much efforts to reveal cosmology on the brane have been
done in the context of five-dimensional spacetime, especially
after the stimulating proposals by Randall and Sundrum (RS)
\cite{Randall:1999ee,Randall:1999vf}. In this model, a
five-dimensional realization of the Horava-Witten solution
\cite{Horava:1995qa}, the hierarchy problem can be solved by
introducing an appropriated exponential warp factor in the metric.
The various properties and characteristics of the RS model have
been extensively analyzed: the cosmology framework
\cite{Shiromizu:1999wj,Garriga:1999yh,
Langlois:2002bb,Maartens:2003tw,Brax:2004xh}, the low energy
effective theory
\cite{Mukohyama:2001ks,Wiseman:2002nn,Langlois:2002hz,Kanno:2002ia,Kanno:2002iaa,Shiromizu:2002qr,Brax:2003vf,Palma:2004fh,
Kobayashi:2006jw,CottaRamusino:2006iu,Fujii:2007fi,Arroja:2007ss},
black hole physics
\cite{Chamblin:1999by,Emparan:1999wa,Chamblin:2000ra,Shiromizu:2000pg,Tamaki:2003bq,Kudoh:2003xz},
the Lorentz violation
\cite{Csaki:2000dm,Stoica:2001qe,Libanov:2005yf,Libanov:2005nv,
Bertolami:2006bf,Ahmadi:2006cr,Ahmadi:2007tr,Koroteev:2009xd,Farakos:2009ka,Farakos:2009ui,Arianto:2009wc},
etc. However, the RS model with codimension one brane world is
insufficient to reconcile a higher-dimensional theory with the
observed four-dimensional spacetime as suggested by string theory.

Recently, the hybrid construction of the Kaluza-Klein and brane
world compactifications, i.e., a Kaluza-Klein compactifications on
the brane has been investigated
\cite{Kanti:2001vb,Charmousis:2004zd,CuadrosMelgar:2005ex,Papantonopoulos:2006uj,Chatillon:2006vw,Yamauchi:2007wm,Kanno:2007wj,Minamitsuji:2008mn}.
Such a way of construction is called Kaluza-Klein brane world. A
basic equation for the study of Kaluza-Klein brane worlds in which
some dimensions on the brane are compactified or for a
regularization scheme for a higher codimension brane world was
derived by Yamauchi and Sasaki \cite{Yamauchi:2007wm}. To analyzes
the Kaluza-Klein cosmology some authors have used the
Shiromizu-Maeda-Sasaki equation \cite{Shiromizu:1999wj} or solving
the bulk geometry. However, it difficult to solve the bulk
geometry in most cases.

In this paper, our main purpose is to study a low energy two brane
cosmological models in higher-dimensional spacetime. We generalize
the case four-dimensional two brane models to $(4+n)$-dimensional
two brane models where $n$ represents internal dimensions of the
brane. We derive the effective equations of motion for
higher-dimensional two brane model using a low energy expansion
method \cite{Kanno:2002ia}. This perturbative method solves the
full $(5+n)$-dimensional equations of motion using an
approximation and after imposing the junction conditions, one
obtains the $(4+n)$-dimensional effective equations of motion. The
effective equations can be solved without knowing the bulk
geometry. Based on this theory we discuss the cosmology two brane
models at low energy. We study in both the static and the
non-static internal dimensions.

This paper is organized as follows. In section
\ref{sec:effective-eq}, we study a higher braneworld model in a
$(5 + n)$-dimensional spacetime bulk with a cosmological constant.
We solve the $(5 + n)$-dimensional Einstein equations at the low
energy using the gradient expansion approximation. We see the
effective theory is described by the $(4+n)$-dimensional
quasi-scalar-tensor gravity with a specific coupling function. In
section \ref{sec:cosmology}, the Kaluza-Klein two brane worlds
cosmology are presented. We derive the effective Friedmann
equations both in the static and non-static internal dimensions.
Section \ref{sec:conclusion} is devoted to the conclusions. In
Appendix \ref{appendix}, we present detailed calculations.

\section{\label{sec:effective-eq}Low energy effective theory for higher-dimensional two brane worlds}
In this section, we derive the low energy effective theory for
higher-dimensional two branes system solving the bulk geometry
formally in the gradient expansion approximation developed by
Kanno and Soda \cite{Kanno:2002ia} (see also
\cite{Kanno:2002iaa}). We consider that the two branes represent a
($4 + n$)-dimensional spacetime embedded in a ($5+n$)-dimensional
spacetime. We assume that there is no matter in the bulk and the
energy-momentum tensor of the bulk is proportional to the
($5+n$)-dimensional cosmological constant, $-2\Lambda_{5+n} =
{(4+n)(3+n)/ l^2}$. Then the higher dimensional braneworld model
is described by the action
\begin{eqnarray}
    S &=& {1\over 2\kappa^2}\int d^{5+n} x \sqrt{-g} \left[{\cal R} + {(4+n)(3+n)\over l^2} \right]\nonumber\\
    &&-\sum_{i=A,B}  \int d^{4+n} x \sqrt{-g^{i{\rm brane}}}\left(\sigma_i-{\cal L}_{\rm matter}^i\right) \ ,
\label{eq:action}
\end{eqnarray}
where ${\cal R}$, $g^{i{\rm brane}}_{\mu\nu}$, $l$ and $\kappa^2 $
are the $(5+n)$-dimensional scalar curvature, the induced metric
on branes, the scale of the bulk curvature radius and the
gravitational constant in $(5+n)$-dimensions, respectively.
Because we will consider the matter terms in (\ref{eq:action}),
the branes will not in general be flat. Consequently we cannot put
both branes at $y = 0$ and $y = l$ and use Gaussian normal
coordinates. Therefore, we use the following coordinate system to
describe the geometry of the brane model,
\begin{equation}
ds^2 = e^{2\phi (y, x^{\mu})} dy^2 + g_{\mu\nu} (y,x^{\mu} )
dx^{\mu} dx^{\nu}  \ . \label{eq:d+1-dim-metric}
\end{equation}
The proper distance between $A$-brane and $B$-brane with fixed $x$
coordinates can be written as
\begin{equation}
    d(x) = \int_0^l e^{\phi (y,x)} dy  \ .
    \label{eq:distande}
\end{equation}
The extrinsic curvature is defined as
\begin{equation}
    K_{\mu\nu} =-{1\over 2}\frac{\partial}{\partial y}g_{\mu\nu} \equiv -{1\over 2}g_{\mu\nu,y}  \ .
    \label{eq:ext-curv}
\end{equation}

In the coordinate system (\ref{eq:d+1-dim-metric}) and using the
extrinsic curvature (\ref{eq:ext-curv}), we can write down the
components of the Einstein equations in ($5+n$)-dimensions as
\begin{widetext}
\begin{eqnarray}
    {}^{(5+n)}G^\mu_{\ \nu } &=& G^\mu_{\ \nu } + e^{-\phi}\left(e^{-\phi}K^\mu_{\ \nu }-\delta^\mu_{\ \nu }e^{-\phi}K\right)_{,y}
                                - (e^{-\phi}K)(e^{-\phi}K^\mu_{\ \nu })\nonumber\\
                            &&+{1\over 2}\delta^\mu_{\ \nu }\left[(e^{-\phi}K)(e^{-\phi}K)
                            + (e^{-\phi}K^{\alpha\beta})(e^{-\phi}K_{\alpha\beta})\right]- \nabla^\alpha \nabla_\alpha \phi -\nabla^\alpha \phi \nabla_\alpha \phi
                            + \delta^\mu_{\ \nu }(\nabla^\alpha \nabla_\alpha \phi + \nabla^\alpha \phi \nabla_\alpha \phi)\nonumber \\
                            &=& {(4+n)(3+n)\over 2 l^2}\delta^\mu_{\ \nu } +\kappa^2 \left(-\sigma^A \delta^\mu_{\ \nu }
                            + T^{A\mu}_{\quad \nu} \right)e^{-\phi} \delta(y) +\kappa^2 \left(-\sigma^B \delta^\mu_{\ \nu }
                            + \tilde{T}^{B\mu}_{\quad \nu} \right)e^{-\phi} \delta(y-l)   \ ,
    \label{eq:compt-munu}\\
    {}^{(5+n)}G^y_{\ y } &=& -{1\over 2} R + {1\over 2} (e^{-\phi} K)(e^{-\phi} K)
                            -{1\over 2} (e^{-\phi} K^{\alpha  \beta})(e^{-\phi} K_{\alpha  \beta} ) = {(4+n)(3+n)\over 2 l^2}  \ ,
    \label{eq:compt-yy} \\
    {}^{(5+n)}G^y_{\ \mu } &=& - \nabla_\nu (e^{-\phi} K_{\mu }^{\  \nu}) + \nabla_\mu (e^{-\phi} K )   =  0   \ ,
    \label{eq:compt-ymu}
\end{eqnarray}
\end{widetext}
where $G^\mu_{\ \nu }=R^\mu_{\ \nu }-\delta^\mu_{\ \nu }R/2$ is
the $(4+n)$-dimensional Einstein tensor and $\nabla_\mu $ denotes
the covariant derivative with respect to the metric $g_{\mu\nu}$.
$T^{\mu}_{\ \nu}$ is the energy momentum tensor of the brane
matter other than the tension. The junction conditions are
obtained by collecting together the terms in field equations which
contain a $\delta$-function, then we obtain
\begin{eqnarray}
    e^{-\phi} \left[ K^\mu_\nu - \delta^\mu_\nu K \right] |_{y=0}
            &=& {\kappa^2 \over 2}
            \left( -\sigma_A \delta^\mu_\nu
            + T^{A\mu}_{\quad\nu} \right) \ ,
            \label{JC-k:A}  \\
    e^{-\phi} \left[ K^\mu_\nu - \delta^\mu_\nu K \right] |_{y=l}
            &=& -{\kappa^2 \over 2}
            \left( -\sigma_B \delta^\mu_\nu
            + \tilde{T}^{B\mu}_{\quad\nu} \right) \ ,
            \label{JC-k:B}
\end{eqnarray}
where $K^\mu_\nu=g^{\mu\alpha}K_{\alpha\nu}$. Note that the
junction conditions constrain the induced metrics on both branes,
they naturally give rise to the effective equations of motion for
the gravity on the branes. In order to solve the bulk field
equations, we use the gradient expansion scheme. The basic idea of
the approximation is the assumption that the energy density of
matter $\rho$ on the brane is smaller than the brane tension
$\sigma$. Equivalently, the bulk curvature scale $l$ is much
smaller than the characteristic length scale of the curvature $L$
on the brane. Then, the small expansion parameter is given by
$\epsilon = \left({l/ L}\right)^2 \ll 1$. This allows us to expand
the metric in perturbative series starting from the induced metric
on the $A$-brane $h_{\mu\nu}$ as the first term
\begin{equation}
     g_{\mu\nu} (y,x^\mu ) =  a^2 (y) \left[ h_{\mu\nu} (x^\mu) + {}^{(1)}g_{\mu\nu} (y,x^\mu)
       + \cdots  \right]  \ ,
\end{equation}
where the boundary conditions on the $A$-brane are given by
\begin{equation}
{}^{(i)}g_{\mu\nu} (y=0 ,x^\mu ) = \left\{\begin{array}{ll}
 h_{\mu\nu}(x^\mu )  & :i=0, \\
0 & :i=1, 2, 3, \ldots
\end{array}
\right.
\end{equation}
For the extrinsic curvature tensor we expand it as
\begin{equation}
   K^\mu_{\ \nu}= {}^{(0)}K^\mu_{\ \nu}+{}^{(1)}K^\mu_{\ \nu}+{}^{(2)}K^\mu_{\ \nu}+ \cdots  \ ,
\end{equation}
where ${}^{(i)}K^\mu_{\ \nu}={\cal O}(\epsilon^i)$.

Applying the above scheme (see Appendix \ref{appendix} for more
detailed), we write down the $(4+n)$-dimensional effective
Einstein equations on the branes in closed form, subject to the
low energy expansion as follows
\begin{eqnarray}
    && G^\mu_{\ \nu} (h) = {(2+n)\kappa^2 \over 2l} T^{A\mu}_{\quad \nu} - {(2+n) \over l}\chi^\mu_{\ \nu} \ ,
    \label{eq:loweq A-brane}\\
    && G^\mu_{\ \nu} (f)  = -{(2+n)\kappa^2 \over 2l} T^{B\mu}_{\quad \nu} - {(2+n) \over l}{\chi^\mu_{\ \nu}\over \Omega^{4+n}}\ ,
    \label{eq:loweq B-brane}
\end{eqnarray}
where the $A$-brane metric is defined as $h_{\mu\nu}\equiv
g^{A-brane}_{\mu\nu}$, while the $B$-brane metric is
$f_{\mu\nu}\equiv g^{B-brane}_{\mu\nu}$. A conformal factor
$\Omega$ relates the metric on the $A$-brane to that on the
$B$-brane, $g^{B-brane}_{\mu\nu}= \Omega^2 g^{A-brane}_{\mu\nu}$.
The terms proportional to $\chi^\mu_{\ \nu}$ are
$(5+n)$-dimensional Weyl tensor contributions, which describe the
non-local $(5+n)$-dimensional effect.

\subsection{Effective theory on $A$-brane }
Eliminating $\chi^\mu_{\ \nu}$ from equations (\ref{eq:loweq
A-brane}) and (\ref{eq:loweq B-brane}), the $(4+n)$-dimensional
field equations on the $A$-brane can be written as
\begin{eqnarray}
   G^\mu_{\ \nu} (h) &=&{(2+n)\kappa^2\over 2l}{ 1 \over \Psi }\left[ T^{A\mu}_{\quad \nu}
                        +(1-\Psi ) T^{B\mu}_{\quad
                        \nu}\right]\nonumber\\
                        &&+{ 1 \over \Psi } \left(  \Psi^{|\mu}_{\ |\nu}
                        -\delta^\mu_\nu  \Psi^{|\alpha}_{\ |\alpha}\right)\nonumber\\
                        &&+{\omega_A \over \Psi^2}\left( \Psi^{|\mu}  \Psi_{|\nu}
                        - {1\over 2} \delta^\mu_\nu  \Psi^{|\alpha} \Psi_{|\alpha}\right)
                         \ ,
   \label{eq:STG-A-brane-1}
\end{eqnarray}
where $|$ denotes the covariant derivative with respect to the
$A$-brane metric $h_{\mu\nu}$ and the new scalar field $\Psi = 1-
\Omega^{2+n}$. The coupling function $\omega_A$ is defined as
\begin{equation}
    \omega_A (\Psi ) \equiv {3+n \over 2+n} {\Psi \over 1-\Psi }  \ .
    \label{eq:coupling}
\end{equation}
We can also determine $\chi^{\mu}_{\nu}$ by eliminating
$G^{\mu}_{\nu}$ from equations (\ref{eq:loweq A-brane}) and
(\ref{eq:loweq B-brane}). Then, we obtain
\begin{eqnarray}
    {(2+n)\over l}\chi^{\mu}_{\ \nu} &=& -{(2+n)\kappa^2\over 2l}{ (1-\Psi) \over \Psi}
                        \left( T^{A\mu}_{\quad \nu} + T^{B\mu}_{\quad \nu} \right)\nonumber\\
                         &&- { 1 \over \Psi } \left(  \Psi^{|\mu}_{\ |\nu}
                        -\delta^\mu_\nu  \Psi^{|\alpha}_{\ |\alpha} \right)\nonumber\\
                        &&+{\omega_A \over \Psi^2} \left( \Psi^{|\mu}  \Psi_{|\nu}
                        - {1\over 2} \delta^\mu_\nu  \Psi^{|\alpha} \Psi_{|\alpha}\right)      \ .
        \label{eq:chi}
\end{eqnarray}
Note that $\chi^{\mu}_{\ \nu}$ is expressed through the quantities
on the branes, $\chi^{\mu}_{\ \nu}=\chi^{\mu}_{\ \nu}(x^\mu)$.
Since $\chi^{\mu}_{\ \nu}$ is traceless, equation (\ref{eq:chi})
leads to an equation of motion for the scalar field $\Psi$,
\begin{eqnarray}
   \Psi^{|\mu}_{\ |\mu} &=& {1\over (3+n) +(2+n)\omega_A}\left[{(2+n)\kappa^2\over 2l}(T^A +T^B)\right.\nonumber\\
                        &&\left.-{d\omega_A \over d\Psi} \Psi^{|\mu}\Psi_{|\mu}\right] \ ,
   \label{eq:STG-A-brane-2}
\end{eqnarray}
where we have taken Eq.~(\ref{eq:coupling}) into account. The
conservation laws for $A$-brane and $B$-brane matter with respect
to the $A$-brane metric $h_{\mu\nu}$ are given by
\begin{equation}
    T^{A\mu}_{\quad \nu |\mu } =0 \ , \quad
    T^{B\mu}_{\quad \nu |\mu } = {\Psi_{|\mu} \over 1-\Psi } T^{B\mu}_{\quad\ \nu}
    -{1\over (2+n)}{\Psi_{|\nu} \over 1-\Psi}T^{B} \ .
    \label{eq:conserve-EM-A-brane}
\end{equation}
One can see that equations (\ref{eq:STG-A-brane-1}) and
(\ref{eq:conserve-EM-A-brane}) do not include the term
$\chi^{\mu}_{\ \nu}$, but they include the energy momentum tensor
of the $B$-brane. For this reason Kanno and Soda called this
theory "quasi-scalar-tensor" gravity.

The effective action on $A$-brane can be derived from the original
$(5+n)$-dimensional action by substituting the solution of the
equations of motion in the bulk and integrating out over the bulk
coordinate. Up to the first order, we obtain the effective action
for $A$-brane as,
\begin{eqnarray}
   && S_A  = {l \over (2+n) \kappa^2} \int d^{4+n} x \sqrt{-h}
            \left[ \Psi R(h) - {\omega_A \over \Psi}
            \Psi^{|\alpha} \Psi_{|\alpha} \right]\nonumber\\
            &&+ \int d^{4+n} x \sqrt{-h} {\cal L}^A+ \int d^{4+n} x \sqrt{-h} \left(1-\Psi \right)^{4+n\over 2+n} {\cal L}^B  \ .
\label{action:d-dim-A-brane}
\end{eqnarray}
Notice that the action (\ref{action:d-dim-A-brane}) represents the
action of the general $(4+n)$-dimensional scalar-tensor theory
with a specific form of the coupling function (\ref{eq:coupling})
and an extra matter term from the $B$-brane.

\subsection{Effective theory on $B$-brane }
To obtain the effective equations of motion on the $B$-brane, we
simply reverse the role of the $A$-brane and that of the
$B$-brane. Solving equation (\ref{eq:loweq B-brane}) for $G^\mu_{\
\nu} (f)$, the $(4+n)$-dimensional field equations on the
$B$-brane can be written as
\begin{eqnarray}
   G^\mu_{\ \nu} (f) &=&{(2+n)\kappa^2 \over 2l}{1\over \Phi} \left[T^{B\mu}_{\quad \nu}
                            +(1+\Phi ) T^{A\mu}_{\quad
                            \nu}\right]\nonumber\\
                            &&+{ 1 \over \Phi } \left( \Phi^{;\mu}_{\ ;\nu}
                            -\delta^\mu_\nu  \Phi^{;\alpha}_{\;\alpha}\right)\nonumber\\
                            && +{\omega_B \over \Phi^2} \left( \Phi^{;\mu}  \Phi_{;\nu}
                            - {1\over 2} \delta^\mu_\nu  \Phi^{;\alpha} \Phi_{;\alpha}\right)   \ ,
   \label{eq:STG-B-brane-1}
\end{eqnarray}
where $;$ denotes the covariant derivative with respect to the
$B$-brane metric $f_{\mu\nu}$ and $\Phi = \Omega^{-(2+n)}-1$.
Here, the coupling function $\omega_B$ is defined as
\begin{equation}
    \omega_B (\Phi ) = -{3+n \over 2+n} {\Phi \over 1+ \Phi }  \ .
    \label{eq:coupling-B}
\end{equation}
The equations of motion for the scalar field $\Phi$ becomes
\begin{eqnarray}
    \Phi^{;\mu}_{\ ;\mu} &=& {1\over (3+n)+(2+n)\omega_B}\left[{(2+n)\kappa^2\over 2l}(T^A +T^B)\right.\nonumber\\
    &&\left. -{d\omega_B \over d\Phi} \Phi^{;\mu}\Phi_{;\mu}\right] \ .
   \label{eq:STG-B-brane-1}
\end{eqnarray}
The conservation laws of the $A$-brane and $B$-brane matter with
respect to the $B$-brane metric $f_{\mu\nu}$ are as follows
\begin{equation}
    T^{A\mu}_{\quad \nu ;\mu } ={\Phi_{;\mu} \over 1+\Phi }
        T^{A\mu}_{\quad \nu}
        -{1\over (2+n)}{\Phi_{;\nu} \over 1+\Phi}T^{A} \ , \quad
    T^{B\mu}_{\quad \nu ;\mu } = 0 \ .
        \label{eq:conserve-EM}
\end{equation}
Finally, the corresponding effective action for $B$-brane is
\begin{eqnarray}
    &&S_B  = {l \over (2+n) \kappa^2} \int d^{4+n} x \sqrt{-f}
            \left[ \Phi R (f)- {\omega_B \over \Phi}
            \Phi^{;\alpha} \Phi_{;\alpha} \right]\nonumber\\
            &&+ \int d^{4+n} x \sqrt{-f} {\cal L}^B+ \int d^{4+n} x \sqrt{-f} \left(1+\Phi \right)^{4+n\over 2+n} {\cal L}^A  \ .
            \label{action:d-dim-B-brane}
\end{eqnarray}

In the derivation of equations of motion above we first to know
the dynamics on one brane. Then we know the gravity on the other
branes. Therefore, the dynamics on both branes are not
independent. The transformation rules for scalar radion and the
metric in $(4+n)$-dimensions are given by
\begin{eqnarray}
    \Phi &=& {\Psi \over 1-\Psi}  \ , \\
    g^{B-{\rm brane}}_{\mu\nu} &=& (1-\Psi )^{2\over (2+n)}\times \nonumber\\
    &&\times\left[ h_{\mu\nu}
    + g^{(1)}_{\mu\nu}\left(h_{\mu\nu} , \Psi, T^A_{\mu\nu } ,  T^B_{\mu\nu } , \ y=l \right) \right] \ .
\end{eqnarray}
The bulk metric is determined if we know the energy momentum
tensors on both branes, the induced metric on $A$-brane, and the
scalar field $\Psi$. Since $(4+n)$-dimensional fields allow us to
construct the $(5+n)$-dimensional bulk geometry, the
quasi-scalar-tensor theory works as a holographic at low energy.

In the following section, for the realization at the first order
expansion, we study the cosmological consequences of the model. We
solve the effective equations without knowing the bulk geometry.
Then, we can determine the Friedman equation on the brane. Here we
focus on the positive tension brane, $A$-brane.


\section{\label{sec:cosmology} Kaluza-Klein two brane worlds cosmology at low energy}

\subsection{Effective Friedmann equation}

In this section, we discuss the cosmological consequences of the
higher-dimensional brane worlds. We take the induced metric on
$A$-brane of the form
\begin{eqnarray}
    ds^2 = -dt^2 + a^2(t)\delta_{ij} dx^i dx^j + b^2(t)\delta_{\alpha\beta} dz^\alpha dz^\beta \ ,
    \label{metric-cos}
\end{eqnarray}
where $\delta_{ij}$ represents the metric of three-dimensional
ordinary spaces with the spatial coordinates $x^i$ ($i=1,2,3$),
while $\delta_{\alpha\beta}$ represents the metric of
$n$-dimensional compact spaces with the coordinates $z^\alpha$
($\alpha = 1, \ldots, n$). The scale factor $b$ denotes the size
of the internal dimensions, while the scale factor $a$ is the
usual scale factor for the external space. We choose the energy
momentum tensors of the $A$-brane and $B$-brane of the following
form
\begin{eqnarray}
     T^{A}_{\mu\nu} &=& (\rho_A, P_A a^2\delta_{ij}, Q_A b^2\delta_{\alpha\beta})\ ,
     \label{EM-A-brane}\\
     T^{B}_{\mu \nu} &=& \Omega^2(\rho_B, P_B a^2\delta_{ij}, Q_B b^2\delta_{\alpha\beta})\ ,
     \label{EM-B-brane}
\end{eqnarray}
where $\rho_i$ is the energy density, $P_i$ the external pressure
and $Q_i$ the internal pressure, $i=A,B$. The $\Omega^2$ factor
results from the fact that the $B$-brane metric is  $f_{\mu\nu}=
\Omega^2h_{\mu\nu}$. The symmetries imply that $\Psi$ only depends
on time.

Using the metric (\ref{metric-cos}) and the energy momentum
tensors (\ref{EM-A-brane}), (\ref{EM-B-brane}) in the effective
Einstein equations (\ref{eq:STG-A-brane-1}), one finds
\begin{widetext}
\begin{eqnarray}
    && 3H^2_a + 3nH_a H_b +{n(n-1)\over 2}H^2_b = {8\pi G\over \Psi}\left[\rho_A + \rho_B(1-\Psi)^{4+n\over 2+n} \right]
     + {1\over \Psi} \left[\frac{(n+3)}{2(n+2)}{\dot{\Psi}^2\over (1-\Psi)} -3H_a\dot{\Psi} -nH_b\dot{\Psi}\right] \ ,
     \label{Fr-1}\\
    && -2\dot{H}_a - 3H^2_a - 2nH_a H_b - n \dot{H}_b -{n(n+1)\over 2}H^2_b =
     {8\pi G\over \Psi}\left[P_A + P_B(1-\Psi)^{4+n\over 2+n}  \right] \nonumber\\
      &&\qquad\qquad\qquad\qquad\qquad\qquad\qquad\qquad\qquad\qquad\qquad
      + {1\over \Psi} \left[\ddot{\Psi} +\frac{(n+3)}{2(n+2)}{\dot{\Psi}^2\over (1-\Psi)} +2H_a\dot{\Psi} +nH_b \dot{\Psi} \right] \ ,
     \label{Fr-2}\\
    && -3\dot{H}_a - 6H^2_a - 3(n-1)H_a H_b - (n-1) \dot{H}_b -{n(n-1)\over 2}H^2_b =
     {8\pi G\over \Psi}\left[Q_A + Q_B(1-\Psi)^{4+n\over 2+n}  \right]  \nonumber\\
     &&\qquad\qquad\qquad\qquad\qquad\qquad\qquad\qquad\qquad\qquad\qquad+ {1\over \Psi} \left[\ddot{\Psi}
     +\frac{(n+3)}{2(n+2)}{\dot{\Psi}^2\over (1-\Psi)} +3H_a\dot{\Psi} +(n-1)H_b \dot{\Psi} \right] \ ,
     \label{Fr-3}
\end{eqnarray}
\end{widetext}
where we have defined the Hubble parameters $H_a=\dot{a}/a$ and
$H_b=\dot{b}/b$ and
\begin{equation}
8\pi G = {(2+n)\kappa^2 \over 2l}.
\end{equation}
In the case $n=0$, the above equations reduce to five-dimensional
brane world. For $n=0$, $\Psi=1$, $\dot{\Psi}=0$, the above
equations reduce to the general relativistic FLRW equations with
barotropic perfect fluid.

The equation of motion for the scalar field $\Psi$ is
\begin{eqnarray}
     \ddot{\Psi} &=&  {8\pi G\over (3+n)}\left[\left(\rho_A-3P_A-nQ_A\right)(1-\Psi)\right.\nonumber\\
     &&\left.+ (\rho_B-3P_B-nQ_B)\left(1-\Psi\right)^{4+n\over 2+n}\right]\nonumber\\
     && - {1\over 2}{\dot{\Psi^2}\over (1-\Psi)} - 3H_a\dot{\Psi}- nH_b\dot{\Psi}\ .
     \label{eos-radion-cos}
\end{eqnarray}
In addition, the conservation laws for the matter with respect to
the $A$-brane metric (\ref{eq:conserve-EM-A-brane}) are given by
\begin{eqnarray}
    & & \dot{\rho}_A + 3 H_a(\rho_A + P_A) +n H_b (\rho_A + Q_A) = 0\ ,
     \label{eq:conserve-EM-A-brane-matt-A}\\
    & & \dot{\rho}_B + 3 H_a(\rho_B + P_B) +n H_b (\rho_B + Q_B) = \nonumber\\
    & &\qquad\qquad\quad\frac{3(\rho_B + P_B)+ n(\rho_B+Q_B)}{2+n}\frac{\dot{\Psi}}{1-\Psi}\ .
     \label{eq:conserve-EM-A-brane-matt-B}
\end{eqnarray}
Substituting equation (\ref{eos-radion-cos}) into equations
(\ref{Fr-2}) and (\ref{Fr-3}), respectively, and assuming the
matter distribution on the branes are given by the equations of
state $P_i=w_i\rho_i$ and $Q_i=v_i\rho_i$ ($i=A, B$). Equations
(\ref{Fr-2}) and (\ref{Fr-3}) reduce to
\begin{widetext}
\begin{eqnarray}
    && -2\dot{H}_a - 3H^2_a - 2nH_a H_b - n \dot{H}_b -{n(n+1)\over 2}H^2_b + H_a{\dot{\Psi}\over \Psi}=
     {8\pi G\over \Psi}\left[w_A \rho_A + {(1-3w_A-nv_A)\over (3+n)}\rho_A (1-\Psi)\right.\nonumber\\
     &&\qquad\qquad\qquad\qquad\qquad\qquad\qquad\qquad\qquad\qquad\qquad\left.
     + {(1+nw_B-nv_B)\over (3+n)}\rho_B (1-\Psi)^{4+n\over 2+n}\right]
     + \frac{1}{2(n+2)}{\dot{\Psi}^2\over \Psi(1-\Psi)}  \ ,
     \label{Fr-2-reduce}\\
   &&  -3\dot{H}_a - 6H^2_a - 3(n-1)H_a H_b - (n-1) \dot{H}_b -{n(n-1)\over 2}H^2_b+H_b{\dot{\Psi}\over \Psi} =
     {8\pi G\over \Psi}\left[v_A \rho_A + {(1-3w_A-nv_A)\over (3+n)}\rho_A (1-\Psi)\right.\nonumber\\
     &&\qquad\qquad\qquad\qquad\qquad\qquad\qquad\qquad\qquad\qquad\qquad\left.
     + {(1-3w_B+3v_B)\over (3+n)}\rho_B (1-\Psi)^{4+n\over 2+n}\right] + \frac{1}{2(n+2)}{\dot{\Psi}^2\over \Psi(1-\Psi)} \ .
     \label{Fr-3-reduce}
\end{eqnarray}
\end{widetext}
From equations (\ref{Fr-1}), (\ref{Fr-2-reduce}), and
(\ref{Fr-3-reduce}), we eliminate $\dot{\Psi}^2$ term to obtain
\begin{widetext}
\begin{eqnarray}
    &&2\dot{H}_a +{3(4+n)\over (3+n)}H^2_a + {n(9+2n)\over (3+n)}H_aH_b+ n\dot{H}_b + {n(n^2+5n+2)\over 2(3+n)}H^2_b
    - {n\over (3+n)}(H_a-H_b){\dot{\Psi}\over \Psi} =\nonumber\\
   && {8\pi G\over \Psi} \left[ {(1-(3+n)w_A)\rho_A\over (3+n)} -{(1-3w_A-nv_A)\rho_A(1-\Psi)\over (3+n)}
    - {n(w_B - v_B)\over (3+n)}\rho_B(1-\Psi)^{4+n\over 2+n}\right]\ ,
    \label{Fr-1-2-reduce}\\
   & & \dot{H}_a +3H^2_a + (n-3)H_aH_b - \dot{H}_b - nH^2_b+ (H_a-H_b){\dot{\Psi}\over
   \Psi}= {8\pi G\over \Psi} \left[(w_A-v_A)\rho_A +(w_B-v_B)\rho_B (1-\Psi)^{4+n\over 2+n} \right]\ .
    \label{Fr-2-3-reduce}
\end{eqnarray}
\end{widetext}
Combining equations (\ref{Fr-1-2-reduce}) and
(\ref{Fr-2-3-reduce}) we get the dynamical equation for Hubble
parameters in $(4+n)$-dimensions,
\begin{eqnarray}
    &&\dot{H}_a +2H^2_a +nH_aH_b + {n(1+n)\over 6}H_b^2 + {n\over 3}\dot{H}_b\nonumber\\
    && =  {8\pi G\over 3}{(1-3w_A - nv_A)\over (2+n)}\rho_A \ .
    \label{Fr-dyn-H}
\end{eqnarray}
The conservation laws reduce to
\begin{eqnarray}
    && \dot{\rho}_A + 3 H_a(1 + w_A)\rho_A +n H_b (1 + v_A)\rho_A = 0\ ,
     \label{eq:conserve-EM-A-brane-matt-A-re}\\
     &&\dot{\rho}_B + 3 H_a(1 + w_B)\rho_B +n H_b (1 + v_B)\rho_B\nonumber\\
    &&= \frac{[3(1 + w_B)+ n(1+v_B)]\rho_B}{2+n}\frac{\dot{\Psi}}{1-\Psi}\ .
     \label{eq:conserve-EM-A-brane-matt-B-re}
\end{eqnarray}
In general, equation (\ref{Fr-dyn-H}) is a second order
differential equation for scale factor $a(t)$ and $b(t)$. In the
case 4-dimensional braneworld ($n=0$), equation (\ref{Fr-dyn-H})
can be solved analytically, and this results in the Friedmann
equation on the brane with the dark radiation term as an
integration constant. In our case equation (\ref{Fr-dyn-H}) cannot
be integrated analytically and therefore, the usual form of the
Friedmann equation on the brane cannot be extracted. In the
following two subsections we consider two cases: static and
non-static internal dimensions.

\subsection{Friedmann equation with static internal dimensions}
In the case of static internal extra dimensions, the dynamical of
the $A$-brane is described by the following equations
\begin{widetext}
\begin{eqnarray}
    & & H^2_a + H_a{\dot{\Psi}\over \Psi} -\frac{(n+3)}{6(n+2)}{\dot{\Psi}^2\over \Psi(1-\Psi)}
      = {8\pi G\over 3\Psi}\left[\rho_A + \rho_B(1-\Psi)^{4+n\over 2+n} \right]   \ ,
     \label{Fr-1 static}\\
    & & \dot{H}_a +2H^2_a  =  {8\pi G\over 3}{(1-3w_A - nv_A)\over (2+n)}\rho_A \ ,
    \label{Fr-static-H}\\
    & & \ddot{\Psi} + 3H_a\dot{\Psi} +{1\over 2}{\dot{\Psi^2}\over (1-\Psi)}=  {8\pi G\over
    (3+n)}\left[\left(1-3w_A-nv_A\right)\rho_A(1-\Psi)
    + (1-3w_B-nv_B)\rho_B\left(1-\Psi\right)^{4+n\over 2+n}\right] \ .
     \label{eos-radion-cos-static-1}
\end{eqnarray}
\end{widetext}
Here we have assumed that the compact dimensions are stabilized,
$b(t)=1$ \cite{Kanno:2007wj}. We see that the above equations do
not contain any additional term compared with five-dimensional
brane world cosmology. However, the differences from the usual two
brane models are concealed in the gravitational constant and also
in the form of the constraint equation (\ref{Fr-1 static}).

The conservation laws for the matter with respect to the $A$-brane
metric reduce to
\begin{eqnarray}
     & &\dot{\rho}_A + 3 H_a(1 + w_A)\rho_A = 0\ ,
     \label{eq:conserve-EM-A-static}\\
     &&\dot{\rho}_B + 3 H_a(1+ w_B)\rho_B   =\nonumber\\
     && \frac{3( 1+ w_B)\rho_B+ n(1+v_B)\rho_B}{2+n}\frac{\dot{\Psi}}{1-\Psi}\ ,
     \label{eq:conserve-EM-B-static}
\end{eqnarray}
and we obtain
\begin{eqnarray}
    && \rho_A \propto a^{-3(1 + w_A)} \ , \\
    && \rho_B \propto a^{-3(1 + w_B)} (1-\Psi)^{\frac{3( 1+ w_B)+ n(1+v_B)}{2+n}}\ ,
\end{eqnarray}
A relation between the energy densities on both branes can
obtained by eliminating $a$,
\begin{eqnarray}
     \rho_B \propto \rho_A^{(1 + w_B)\over (1 + w_A)} (1-\Psi)^{\frac{3( 1+ w_B)+ n(1+v_B)}{2+n}}\ ,
\end{eqnarray}

In the case $w_A\neq 1/3$, leaving $v_A$ as a free parameter and
using the matter conservation equation
(\ref{eq:conserve-EM-A-static}) we can write (\ref{Fr-static-H})
as
\begin{eqnarray}
    {d\over dt}\left(a^4 H_a^2 - {8 \pi G\over 3}{2(1-3w_A-nv_A)\over (2+n)(1-3w_A)} a^4 \rho_A\right)=0 \ .
\end{eqnarray}
Then, we obtain an expression for the effective Hubble parameter
on $A$-brane as
\begin{eqnarray}
    H^2_a = {8 \pi G_{eff}\over 3} \rho_A + {{\cal C}\over a^4} \ ,
    \label{sol-H}
\end{eqnarray}
where ${\cal C}$ is is an integration constant which can be
interpreted as dark radiation. We have defined the effective
gravitational constant
\begin{equation}
    G_{eff} ={2(1-3w_A-nv_A)\over (2+n)(1-3w_A)}\ G\ .
    \label{grav-eff-gen}
\end{equation}
For $w_A<1/3$, $nv_A<1-3w_A$ and $w_A>1/3$, $nv_A>1-3w_A$, the
effective gravitational constant becomes positive.

In the case of radiation dominated universe, $w_A=1/3$, we have
\begin{eqnarray}
    \dot{H}_a +2H^2_a  = - {8\pi Gnv_A \over 3(2+n)}\rho_A \ ,
    \label{Fr-static-H-rad}
\end{eqnarray}
Using the matter conservation equation, we can write equation
(\ref{Fr-static-H-rad}) as
\begin{eqnarray}
    {d\over dt} \left(a^4 H_a^2 + {8\pi G\over 3}{2nv_A \over (2+n)}\log
    a\right)=0\ ,
\end{eqnarray}
and giving
\begin{eqnarray}
    H_a^2  =-{8\pi G\over 3}{2nv_A \log a \over (2+n)} \rho_A
    + {K\over a^4 } \ ,
    \label{rad-sol}
\end{eqnarray}
where $K$ is an integration constant which can be redefined as a
sum of the initial value of radiative matter density and initial
value of the dark radiation density ${\cal C}$. Then equation
(\ref{rad-sol}) becomes
\begin{eqnarray}
    H_a^2  ={8\pi G\over 3}\left(1-{2n \over (2+n)}v_A \log {a\over a_{*}} \right) \rho_A
    + {{\cal C}\over a^4 } \ ,
    \label{rad-sol-1}
\end{eqnarray}
where $a_{*}$ is a constant corresponding to the dark radiation
component ${\cal C}$. Defining the effective gravitational
constant
\begin{equation}
    G_{eff} =\left[1-{2n\over (2+n)}v_A\log {a\over a_{*}}\right]G \ ,
    \label{grav-eff-rad}
\end{equation}
then we have the effective Friedmann equation (\ref{sol-H}). As
expected the expression for the effective Friedmann equation on
$A$-brane coincide with the Kaluza-Klein brane world cosmology
with one brane model in the low energy approximation where the
term of quadratic energy density is neglected \cite{Kanno:2007wj}.
In contrast to the usual four-dimensional two-brane model, the
effective gravitational constant depends  on the equation of state
and the external scale factor explicitly, and may becomes positive
or negative.

\subsection{Friedmann equation with non-static internal dimensions}
Let us now consider the case of non-static internal dimensions, in
which the brane world evolves with two scale factors. We take a
simple relation between the scale factors on $A$-brane of the form
\begin{eqnarray}
    b(t)  =  a^{\gamma}(t) \ ,
    \label{dyn-internal}
\end{eqnarray}
where $\gamma$ is a constant. For the internal scale factor $b(t)$
to be small compared to the external scale factor $a(t)$, the
constant $\gamma$ should be negative.

For non-static internal dimensions, the dynamical of $A$-brane is
described by the following equations
\begin{widetext}
\begin{eqnarray}
    & & \left[{6(1 + n\gamma) +n(n-1)\gamma^2\over 2} \right]H^2_a+ (3+n\gamma)H_a{\dot{\Psi}\over \Psi}=
     {8\pi G\over \Psi}\left[\rho_A + \rho_B(1-\Psi)^{4+n\over 2+n} \right]
     + \frac{(n+3)}{2(n+2)}{\dot{\Psi}^2\over \Psi(1-\Psi)}  \ ,
     \label{Fr-1-dyn}\\
    & & \dot{H}_a +{6(2+n\gamma)+n(1+n)\gamma^2 \over 2(3+n\gamma)}H_a^2  = {8\pi G(1-3w_A - nv_A)\over (2+n)(3+n\gamma)}\rho_A \ ,
    \label{Fr-dyn-H-dyn-int}\\
   & & \ddot{\Psi} + (3+ n\gamma)H_a\dot{\Psi}=  {8\pi G\over (3+n)}\left[\left(1-3w_A-nv_A\right)\rho_A(1-\Psi)
     + (1-3w_B-nv_B)\rho_B\left(1-\Psi\right)^{4+n\over 2+n}\right]
     - {1\over 2}{\dot{\Psi^2}\over (1-\Psi)} \ ,
     \label{eos-radion-cos-dyn}
\end{eqnarray}
\end{widetext}
The conservation laws become
\begin{eqnarray}
     &&\dot{\rho}_A + \left[3(1 + w_A) +n\gamma(1 + v_A)\right]H_a\rho_A = 0\ ,
     \label{eq:conserve-EM-A-brane-matt-A-dyn}\\
     &&\dot{\rho}_B + \left[3(1 + w_B) + n\gamma  (1+ v_B)\right]H_a\rho_B
     \nonumber\\
     && =\frac{[3(1 + w_B)+ n(1+v_B)]\rho_B}{2+n}\frac{\dot{\Psi}}{1-\Psi}\ ,
     \label{eq:conserve-EM-A-brane-matt-B-dyn}
\end{eqnarray}
Using the matter conservation equation
(\ref{eq:conserve-EM-A-brane-matt-A-dyn}),
\begin{eqnarray}
    &&\left[4\beta -3(1+w_A)-n\gamma(1+v_A)\right]H_a\rho_A
    \nonumber\\
    &&=\dot{\rho}+ 4\beta H_a\rho_A = {1\over a^{4\beta}}{d\over dt}\left(a^{4\beta}\rho_A\right) \ ,
\end{eqnarray}
and so we can write equation (\ref{Fr-dyn-H-dyn-int}) as
\begin{eqnarray}
    {d\over dt}\left(a^{4\beta}H_a^2-{8\pi G_{eff}\over 3}a^{4\beta}\rho_A\right)=0 \ ,
\end{eqnarray}
where
\begin{eqnarray}
    \beta = {6(2+n\gamma)+n(1+n)\gamma^2 \over 4(3+n\gamma)} \ .
    \label{beta}
\end{eqnarray}
Then the effective Friedmann equation for non-static internal
dimensions on $A$-brane is given by
\begin{eqnarray}
    H^2_a  =  {8\pi G_{eff}\over 3}\rho_A + {C \over a^{4\beta}} \ ,
    \label{fr-nonstatic-1}
\end{eqnarray}
where $C$ is a constant of integration and we have defined the
effective gravitational constant as follows
\begin{equation}
   G_{eff}= {6(1-3w_A-nv_A)G\over (2+n)(3+n\gamma)[4\beta-3(1+w_A)-n\gamma(1+v_A)]} \ .
\end{equation}
Notice that for $n=3$ and non-static internal dimensions, the
setup is symmetric under the exchange of internal and external
pressures ($w_i \leftrightarrow v_i$), and $a(t) \leftrightarrow
b(t)$.

The above results also include the well-known five dimensional
brane world, corresponding to $n=0$ and for which $\beta=1$,
$G_{eff}=G$. For $\gamma=0$ the above results reduce to the static
internal dimensions. If $\gamma=1$, the scale factor $b(t)$ is
related to $a(t)$ as $b(t)=a(t)$, we obtain the Friedmann equation
of the generalized Randall-Sundrum model in $(5+n)$ dimensions
describing a $(4+n)$-dimensional universe.
\begin{eqnarray}
    H^2_a  =  {8\pi G_{eff}\over 3}\rho_A + {C \over a^{4+n}} \ ,
\end{eqnarray}
where the effective gravitational constant is now given by
\begin{eqnarray}
   G_{eff}= {6\over (2+n)(3+n)}\ G \ .
\end{eqnarray}
In the case $n=0$, the above Friedmann equation reduces to usual
Friedmann equation on four-dimensional brane.

Leaving $\beta$ as a free parameter, we can solve equation
(\ref{beta}) for $\gamma$. We obtain
\begin{equation}
    \gamma = - \frac{3-2\beta \pm \sqrt{{4\beta(3+n\beta)-3(4+n)\over n}}}{1+n}\ .
    \label{gamma-para}
\end{equation}
The negative values of $\gamma$ indicate that the internal scale
factor $b(t)$ to be small compared to the external scale factor
$a(t)$. Taking $\beta=1$ such that the second term of Friedmann
equation (\ref{fr-nonstatic-1}) contributes the "dark" radiation,
we have
\begin{equation}
    \gamma = -{2\over 1+n}, \quad \text{or} \quad \gamma = 0 \ ,
    \label{int-dark-rad}
\end{equation}
where $\gamma = 0$ corresponds to the static internal dimensions.
Therefore, the "dark" radiation component in the Friedmann
equation can be also realized in the Kaluza-Klein brane worlds
with non-static internal dimensions.

\subsection{Hubble parameters in conformal frames}

The action on $A$-brane is written in the Jordan frame, for which
the gravitational sector has a non-canonical form. We can,
however, perform a conformal transformation to the Einstein frame:
$\tilde{h}_{\mu\nu}=\Psi^{2/(2+n)} h_{\mu\nu}$. In the Einstein
frame, the metric (\ref{metric-cos}) is
\begin{eqnarray}
    d\tilde{s}^2 &=& \tilde{h}_{\mu\nu} dx^\mu dx^\nu \nonumber\\
    &=& \Psi^{2\over(2+n)} \left[-dt^2 + a^2(t)\delta_{ij} dx^i dx^j + b^2(t)\delta_{\alpha\beta} dz^\alpha dz^\beta\right] \nonumber\\
    &=& -d\tilde{t}^2 + \tilde{a}^2(\tilde{t})\delta_{ij} dx^i dx^j + \tilde{b}^2(\tilde{t})\delta_{\alpha\beta} dz^\alpha dz^\beta\ ,
    \label{metric-cos-Einstein}
\end{eqnarray}
and the Hubble parameters satisfy
\begin{eqnarray}
    \tilde{H}_a - \tilde{H}_b = \Psi^{-{1\over 2+n}}(H_a - H_b) \ ,
\end{eqnarray}
where $\tilde{H}_a=\tilde{a}^{-1}(d\tilde{a}/d\tilde{t})$ and
$\tilde{H}_b=\tilde{b}^{-1}(d\tilde{b}/d\tilde{t})$. One can see
that the static internal dimensions (in the Jordan frame) may
becomes dynamics in the Einstein frame. In this case we have,
\begin{eqnarray}
    \tilde{H}_a - \tilde{H}_b = \Psi^{-{1\over 2+n}}H_a,\quad \tilde{H}_b = {1\over (2+n)\Psi}{d\Psi\over d\tilde{t}} \ .
\end{eqnarray}
In the case $b(t)=a^\gamma(t)$, we have
\begin{eqnarray}
    \tilde{H}_a - \tilde{H}_b =  ( 1- \gamma)\Psi^{-{1\over 2+n}}H_a \ .
\end{eqnarray}
Dynamics of the Hubble parameters $H_a$ and $H_b$ in the Jordan
frame are also dynamics in the Einstein frame.

\section{\label{sec:conclusion} Conclusion}
In this paper we have derived the low energy effective equations
for the higher-dimensional two brane models by using gradient
expansion approximation. As expected, the effective theory is
described by the $(4+n)$-dimensional quasi-scalar-tensor gravity
with a specific coupling function. The presented effective
equations can be used as the basic equations for the
higher-dimensional two brane worlds cosmology, in which some
spatial dimensions on the brane are Kaluza-Klein compactified.

We can see already from the Friedmann equations that the
Kaluza-Klein brane world can be realized at low energies. Due to
their complicated structure the field equations appearing in the
theories are very difficult to solve analytically, we have
restricted our discussions with the special cases: static internal
dimensions and non-static internal dimensions where a relation
between the external and internal scale factors is given by
$b(t)=a^\gamma(t)$. In the static internal dimensions $\gamma=0$,
our results coincide with the Kaluza-Klein brane world cosmology
with one brane model in the low energy approximation where the
term of quadratic energy density is neglected \cite{Kanno:2007wj}.
In the non-static internal dimensions, the induced Friedmann
equation on the brane is modified in the effective gravitational
constant and the term proportional to $a^{-4\beta}$.

Another important result of this work is the dynamics of the
internal Hubble parameter in conformal frames. Both the static and
non-static internal dimensions in the Jordan frame are always
dynamics in the Eintein frame. However, the physical
interpretation and equivalence of these two frames is a problem in
the case of static internal dimensions in the Jordan frame. We
plan to investigate the correspondence between the Jordan and the
Einstein frame description, including the dynamical of scalar
field.

\begin{acknowledgements}
We thank the referees for many helpful suggestions which improved
the presentation of the paper.
\end{acknowledgements}

\appendix

\section{\label{appendix}Detailed calculations}
Let us decompose the extrinsic curvature into the traceless part
and the trace part
\begin{equation}
    e^{-\phi}K_{\mu\nu} = \Sigma_{\mu\nu} + {1\over 4+n} g_{\mu\nu} Q  \ , \quad
    Q = - e^{-\phi}{\partial \over \partial y}\log \sqrt{-g}    \
    ,
    \label{eq:decompose}
\end{equation}
which allows us to write the field equations (\ref{eq:compt-munu})
- (\ref{eq:compt-ymu}) in the bulk as follows
\begin{eqnarray}
    & & e^{-\phi} \Sigma^\mu_{\ \nu , y} - Q \Sigma^\mu_{\ \nu}
            = -\left[ R^\mu_{\ \nu} - {1\over 4+n} \delta^\mu_{\ \nu} R
                \right.\nonumber\\
                &&\left.-\nabla^\mu \nabla_\nu \phi
                -\nabla^\mu \phi  \nabla_\nu \phi\right.\nonumber\\
                &&\left.+{1\over 4+n} \delta^\mu_\nu
                \left( \nabla^\alpha \nabla_\alpha \phi
            +\nabla^\alpha \phi \nabla_\alpha \phi \right)
            \right]      \ ,
            \label{eq:munu-traceless} \\
    & & {3+n \over 4+n} Q^2 - \Sigma^\alpha_{\ \beta} \Sigma^\beta_{\ \alpha}
            = \left[ R \right] + {(4+n)(3+n)\over l^2}   \ ,
            \label{eq:munu-trace} \\
    & & e^{-\phi}  Q_{, y} -{1\over 4+n}Q^2
            - \Sigma^{\alpha \beta} \Sigma_{\alpha \beta}
            = \nabla^\alpha \nabla_\alpha \phi
            + \nabla^\alpha \phi \nabla_\alpha \phi \nonumber\\
            &&- {4+n\over l^2}  \ ,
            \label{eq:yy}  \\
    & & \nabla_\lambda \Sigma_{\mu }^{\ \lambda}
            - {3+n\over 4+n} \nabla_\mu Q = 0   \ .
            \label{eq:ymu}
\end{eqnarray}
The junction conditions determine the dynamics of the induced
metric and provide the effective theory of gravity on the brane
reduced to
\begin{eqnarray}
    \left[ \Sigma^{\mu}_{\nu}
        - {3\over 4} \delta^\mu_\nu Q \right] \Bigg|_{y=0}
            &=& {\kappa^2 \over 2} (-\sigma_A \delta^\mu_\nu
            + T^{A\mu}_{\quad\nu})  \ ,
            \label{JC:A} \\
        \left[ \Sigma^{\mu}_{\nu}
        - {3\over 4} \delta^\mu_\nu Q \right] \Bigg|_{y=l}
            &=& -{\kappa^2 \over 2} (-\sigma_B \delta^\mu_\nu +
            \tilde{T}^{B\mu}_{\quad\nu})    \ .
            \label{JC:B}
\end{eqnarray}
\subsection{Zeroth order}
At zeroth order, the gradient terms and matter on the brane can be
ignored. We find
\begin{eqnarray}
    {}^{(0)}\Sigma^{\mu}_{\nu} =0, \qquad  {}^{(0)}Q={4+n\over l} \ .
\end{eqnarray}
The junction conditions (\ref{JC:A}) and (\ref{JC:B}) yield
\begin{eqnarray}
    \sigma_A= {2(3+n)\over \kappa^2 l}, \qquad  \sigma_B= - {2(3+n)\over \kappa^2 l} \ .
\end{eqnarray}
Using the definition of the extrinsic curvature, we get the zeroth
order metric as
\begin{eqnarray}
    &&ds^2 = e^{2\phi(y,x)}dy^2 + a^2(y,x) h_{\mu\nu}dx^\mu dx^\nu,\\
    && a(y,x) = \exp\left[-{1\over l}\int_0^y dy e^{\phi(y,x)}
    \right]\ ,
\end{eqnarray}
where  the tensor $h_{\mu\nu}$ is  the induced metric on
$A$-brane. To proceed we will assume $\phi(y,x)\equiv \phi(x)$
thus $a(y,x) = \exp\left[-{ye^{\phi(x)}/ l}\right]$.

\subsection{First order}
In the first order, the curvature term that has been ignored in
the zeroth order calculation comes into play. Substituting the
solutions at zeroth order, the field equations
(\ref{eq:munu-traceless}) - (\ref{eq:ymu}) can be written as
follows
\begin{eqnarray}
    & & e^{-\phi} {}^{(1)}\Sigma^\mu_{\ \nu , y} - {4+n \over l}  {}^{(1)}\Sigma^\mu_{\ \nu}
            = -\left[ R^\mu_{\ \nu} - {1\over 4+n} \delta^\mu_{\ \nu} R\right.\nonumber\\
                &&\left.
                -(\nabla^\mu \nabla_\nu \phi+\nabla^\mu \phi  \nabla_\nu \phi)\right.\nonumber\\
                &&\left.
                +{1\over 4+n} \delta^\mu_\nu\left( \nabla^\alpha \nabla_\alpha \phi
            +\nabla^\alpha \phi \nabla_\alpha \phi \right)\right]^{(1)}      \ ,
            \label{eq:munu-traceless-1} \\
    & & {2(3+n)\over l} {}^{(1)}Q =\left[ R \right]^{(1)}    \ ,
            \label{eq:munu-trace-1} \\
    & & e^{-\phi}  {}^{(1)}Q_{, y} -{2\over l}{}^{(1)}Q= \left[ \nabla^\alpha \nabla_\alpha \phi
            + \nabla^\alpha \phi \nabla_\alpha \phi \right]^{(1)}  \ ,
            \label{eq:yy-1}  \\
    & & \nabla_\lambda {}^{(1)}\Sigma_{\mu }^{\ \lambda}
            - {3+n\over 4+n} \nabla_\mu {}^{(1)}Q = 0   \ .
            \label{eq:ymu-1}
\end{eqnarray}
And the junction conditions are given by
\begin{eqnarray}
    \left[ {}^{(1)}\Sigma^{\mu}_{\nu}
        - {3\over 4} \delta^\mu_\nu {}^{(1)}Q \right] \Bigg|_{y=0}
            &=& {\kappa^2 \over 2} T^{A\mu}_{\quad\nu}  \ ,
            \label{JC:A-1} \\
        \left[ {}^{(1)}\Sigma^{\mu}_{\nu}
        - {3\over 4} \delta^\mu_\nu {}^{(1)}Q \right] \Bigg|_{y=l}
            &=& -{\kappa^2 \over 2} \tilde{T}^{B\mu}_{\quad\nu}  \ .
            \label{JC:B-1}
\end{eqnarray}
where the superscript $(1)$ represents the order of the gradient
expansion. Now one can express the Ricci tensor $[R^{\mu}_{\
\nu}(g)]^{(1)}$ in term of the Ricci tensor of the $A$-brane
metric $h_{\mu\nu} \equiv g^{A-brane}_{\mu\nu} $ (denoted by
$R^\mu_{\ \nu} (h)$) and $\phi$;
\begin{eqnarray}
    \left[R^\mu_{\ \nu}(g) \right]^{(1)}
        &=& {1\over a^2}\left[  R^\mu_{\ \nu} (h)
        + {(2+n)ye^{\phi}\over l}  \left(
        \phi^{|\mu}_{\ |\nu}+\phi^{|\mu} \phi_{|\nu}
        \right)\right.\nonumber\\
        &&\left.+ {ye^{\phi}\over l}  \delta^\mu_\nu\left(
        \phi^{|\alpha}_{\ |\alpha}+\phi^{|\alpha} \phi_{|\alpha}\right)\right.\nonumber\\
        &&\left.+ {(2+n)y^2e^{2\phi} \over l^2}  \phi^{|\mu} \phi_{|\nu}\right.\nonumber\\
        &&\left. -  {(2+n)y^2e^{2\phi} \over l^2}
        \delta^\mu_\nu\phi^{|\alpha} \phi_{|\alpha} \right]     \ ,
        \label{Ricci-A-brane-O-1}
\end{eqnarray}
where $|$ denotes the covariant derivative with respect to the
$A$-brane metric  $h_{\mu\nu}$. Taking trace of equation
(\ref{Ricci-A-brane-O-1}) and using equation
(\ref{eq:munu-trace-1}), the trace part of the extrinsic curvature
can be obtained without solving the bulk geometry,
\begin{eqnarray}
    {}^{(1)}Q(y,x)&=& {l\over 2(3+n)a^2}\left[  R (g)\right]\nonumber\\
        &=& {l\over a^2}\left[ {1\over 2(3+n)} R (h)
        + {ye^{\phi}\over l}  \left(
        \phi^{|\alpha}_{\ |\alpha}+\phi^{|\alpha} \phi_{|\alpha} \right)\right.\nonumber\\
        &&\left.
        - {(2+n)y^2e^{2\phi} \over 2l^2}  \phi^{|\alpha} \phi_{|\alpha} \right]     \ ,
        \label{Trace-Ricci-A-brane-O-1}
\end{eqnarray}
The second derivatives of $\phi$ are given by
\begin{equation}
    \left[ \nabla^\mu \nabla_\nu \phi  \right]^{(1)}= {1\over a^2} \left[
        \phi^{|\mu}_{\ |\nu} + 2{ye^{\phi}\over l} \phi^{|\mu}
        \phi_{|\nu} - {ye^{\phi}\over l} \delta^\mu_\nu \phi^{|\alpha}
        \phi_{|\alpha}  \right].
        \label{phi-O-1}
\end{equation}
It is easy to see that the Hamiltonian constraint equation
(\ref{eq:yy-1}) is trivially satisfied now. Then, equation
(\ref{eq:munu-traceless-1}) can be integrated to give
\begin{eqnarray}
    {}^{(1)}\Sigma^\mu_{\ \nu}(y,x) &=&  {l \over  a^2 } \left[
            {1\over (2+n)}\left( R^\mu_{\ \nu}- {1\over 4+n} \delta^\mu_\nu R \right)\right.\nonumber\\
        &&\left.
            + {y e^{\phi} \over l}\left( \phi^{|\mu}_{\ |\nu}
            -{1\over 4+n}\delta^\mu_\nu  \phi^{|\alpha}_{\
            |\alpha}\right)\right.\nonumber\\
            &&\left.+\left( {y^2 e^{2\phi} \over l^2} + {y e^{\phi} \over l}\right)\times\right.\nonumber\\
        &&\left.
            \times\left(  \phi^{|\mu} \phi_{|\nu} -{1\over 4+n}\delta^\mu_\nu
            \phi^{|\alpha}  \phi_{|\alpha} \right)\right] \nonumber\\
        &&+ {\chi^\mu_{\ \nu} (x)\over a^{4+n}}  \ ,
            \label{eq:munu-traceless-1-sol}
\end{eqnarray}
where $\chi^\mu_{\ \nu}(x)$ is an integration constant whose trace
vanishes: $\chi_\mu^\mu =0$, and equation (\ref{eq:ymu-1})
requires that $\chi^{\mu}_{\ \nu|\mu}=0 $.

Substituting Eqs.~(\ref{Trace-Ricci-A-brane-O-1}) and
(\ref{eq:munu-traceless-1-sol}) into the junction condition at the
$A$-brane (\ref{JC:A-1}), we obtain
\begin{equation}
    {l\over (2+n) } G^\mu_{\ \nu} (h) + \chi^\mu_{\ \nu}
        = {\kappa^2 \over 2} T^{A\mu}_{\quad \ \nu} \ ,
        \label{1:einstein-A}
\end{equation}
and the junction condition at the $B$-brane (\ref{JC:B-1}) yields
\begin{eqnarray}
    &&{l\over (2+n) \Omega^2 }  G^\mu_{\ \nu}
        + { l e^{\phi} \over \Omega^2} \left(  \phi^{|\mu}_{\ |\nu}
        -\delta^\mu_\nu  \phi^{|\alpha}_{\
        |\alpha}\right.\nonumber\\
        &&\left.+  \phi^{|\mu}  \phi_{|\nu} -  \delta^\mu_\nu
        \phi^{|\alpha}  \phi_{|\alpha} \right)\nonumber\\
        &&+ {l e^{2\phi} \over \Omega^2 } \left( \phi^{|\mu} \phi_{|\nu}
        + {(1+n)\over 2} \delta^\mu_\nu  \phi^{|\alpha} \phi_{|\alpha}
        \right)\nonumber\\
        &&+ {\chi^\mu_{\ \nu} \over \Omega^{4+n}}
        = -{\kappa^2 \over 2 \Omega^2 } T^{B\mu}_{\quad \nu}  \ ,
        \label{1:equation-B}
\end{eqnarray}
where $\Omega (x) = a(y=l,x)= \exp[-e^{\phi}]$ and the index of
$T^{B\mu}_{\quad \ \nu}$ is the energy momentum tensor with the
index raised by the induced $A$-brane metric $h_{\mu\nu}$, while
$\tilde{T}^{B\mu}_{\quad \ \nu}$ is the one raised by the induced
metric on the $B$-brane, $f_{\mu\nu} \equiv g^{B-brane}_{\mu\nu}$.
Using $f_{\mu\nu} = \Omega^2
h_{\mu\nu}=\exp[-2e^{\phi}]h_{\mu\nu}$, equation
(\ref{1:equation-B}) can be rewritten as
\begin{equation}
    {l\over (2+n) } G^\mu_{\ \nu} (f) + {\chi^\mu_{\ \nu}\over \Omega^{4+n}}
        = - {\kappa^2 \over 2} \tilde{T}^{B\mu}_{\quad \ \nu} \ ,
        \label{1:einstein-B}
\end{equation}

We now solve the metric in the bulk. The definition
(\ref{eq:decompose}) gives
\begin{equation}
    -\frac{e^{-\phi}}{2a^2}h^{\alpha\mu}\frac{\partial}{\partial y}
        {}^{(1)}g_{\alpha\nu} = {}^{(1)}\Sigma^{\mu}_{\ \nu} + \frac{1}{4+n}
        \delta^{\mu}_{\nu} {}^{(1)}Q \ .
        \label{1:k}
\end{equation}
Integrating Eq.~(\ref{1:k}), we obtain the metric in the bulk:
\begin{eqnarray}
    {}^{(1)}g_{\mu\nu}(y,x) &&= -{l^2 \over (2+n) }\left({1\over a^2} -1 \right)\times\nonumber\\
            &&\times \left[ R_{\mu\nu}  - {1\over 2(3+n)} h_{\mu\nu}R\right]\nonumber\\
            &&+ {l^2 \over 2}\left( {1 \over a^2 }-1 -{2y e^{\phi}\over l}{1\over a^2} \right)\times\nonumber\\
            &&\times \left( \phi_{|\mu \nu}  + {1\over 2} h_{\mu\nu}
            \phi^{|\alpha}  \phi_{|\alpha} \right)   \nonumber \\
            && \quad   -{y^2 e^{2\phi} \over a^2}
            \left(  \phi_{|\mu}  \phi_{|\nu}-  {1\over 2} h_{\mu\nu}
            \phi^{|\alpha}  \phi_{|\alpha}  \right)\nonumber\\
            && -{2l \over 4+n}\left({1\over a^{4+n}} -1 \right)   \chi_{\mu\nu} \ ,
            \label{1:metric}
\end{eqnarray}
where we have imposed the boundary condition, ${}^{(1)}g_{\mu\nu}
(y=0, x^\mu ) =0 $. We can use a schematic iteration
\cite{Kanno:2002ia} for the solutions at higher orders.



\begin{thebibliography}{99}

\bibitem{Randall:1999ee}
  L.~Randall and R.~Sundrum,
  Phys.\ Rev.\ Lett.\  {\bf 83}, 3370 (1999)
  [arXiv:hep-ph/9905221].

\bibitem{Randall:1999vf}
  L.~Randall and R.~Sundrum,
  Phys.\ Rev.\ Lett.\  {\bf 83}, 4690 (1999)
  [arXiv:hep-th/9906064].

\bibitem{Horava:1995qa}
  P.~Horava and E.~Witten,
  Nucl.\ Phys.\  B {\bf 460}, 506 (1996)
  [arXiv:hep-th/9510209].

\bibitem{Shiromizu:1999wj}
  T.~Shiromizu, K.~i.~Maeda and M.~Sasaki,
  Phys.\ Rev.\  D {\bf 62}, 024012 (2000)
  [arXiv:gr-qc/9910076].

\bibitem{Garriga:1999yh}
  J.~Garriga and T.~Tanaka,
  Phys.\ Rev.\ Lett.\  {\bf 84}, 2778 (2000)
  [arXiv:hep-th/9911055].

\bibitem{Langlois:2002bb}
  D.~Langlois,
  Prog.\ Theor.\ Phys.\ Suppl.\  {\bf 148}, 181 (2003)
  [arXiv:hep-th/0209261].

\bibitem{Maartens:2003tw}
  R.~Maartens,
  Living Rev.\ Rel.\  {\bf 7}, 7 (2004)
  [arXiv:gr-qc/0312059].

\bibitem{Brax:2004xh}
  P.~Brax, C.~van de Bruck and A.~C.~Davis,
  Rept.\ Prog.\ Phys.\  {\bf 67}, 2183 (2004)
  [arXiv:hep-th/0404011].


\bibitem{Mukohyama:2001ks}
  S.~Mukohyama and L.~Kofman,
  Phys.\ Rev.\  D {\bf 65}, 124025 (2002)
  [arXiv:hep-th/0112115].

\bibitem{Wiseman:2002nn}
  T.~Wiseman,
  Class.\ Quant.\ Grav.\  {\bf 19}, 3083 (2002)
  [arXiv:hep-th/0201127].

\bibitem{Langlois:2002hz}
  D.~Langlois and L.~Sorbo,
  Phys.\ Lett.\  B {\bf 543}, 155 (2002)
  [arXiv:hep-th/0203036].

\bibitem{Kanno:2002iaa}
  S.~Kanno and J.~Soda,
  Phys.\ Rev.\  D {\bf 66}, 043526 (2002)
  [arXiv:hep-th/0205188].

\bibitem{Kanno:2002ia}
  S.~Kanno and J.~Soda,
  Phys.\ Rev.\  D {\bf 66}, 083506 (2002)
  [arXiv:hep-th/0207029].

\bibitem{Shiromizu:2002qr}
  T.~Shiromizu and K.~Koyama,
  Phys.\ Rev.\  D {\bf 67}, 084022 (2003)
  [arXiv:hep-th/0210066].

\bibitem{Brax:2003vf}
  P.~Brax, C.~van de Bruck, A.~C.~Davis and C.~S.~Rhodes,
  arXiv:hep-ph/0309180.

\bibitem{Palma:2004fh}
  G.~A.~Palma and A.~C.~Davis,
  Phys.\ Rev.\  D {\bf 70}, 064021 (2004)
  [arXiv:hep-th/0406091].

\bibitem{Kobayashi:2006jw}
  T.~Kobayashi, T.~Shiromizu and N.~Deruelle,
  Phys.\ Rev.\  D {\bf 74}, 104031 (2006)
  [arXiv:hep-th/0608166].

\bibitem{CottaRamusino:2006iu}
  L.~Cotta-Ramusino and D.~Wands,
  Phys.\ Rev.\  D {\bf 75}, 104001 (2007)
  [arXiv:hep-th/0609092].

\bibitem{Fujii:2007fi}
  S.~Fujii, T.~Kobayashi and T.~Shiromizu,
  Phys.\ Rev.\  D {\bf 76}, 104052 (2007)
  [arXiv:0708.2534 [hep-th]].

\bibitem{Arroja:2007ss}
  F.~Arroja, T.~Kobayashi, K.~Koyama and T.~Shiromizu,
  JCAP {\bf 0712}, 006 (2007)
  [arXiv:0710.2539 [hep-th]].

\bibitem{Chamblin:1999by}
  A.~Chamblin, S.~W.~Hawking and H.~S.~Reall,
  Phys.\ Rev.\  D {\bf 61}, 065007 (2000)
  [arXiv:hep-th/9909205].

\bibitem{Emparan:1999wa}
  R.~Emparan, G.~T.~Horowitz and R.~C.~Myers,
  JHEP {\bf 0001}, 007 (2000)
  [arXiv:hep-th/9911043].

\bibitem{Chamblin:2000ra}
  A.~Chamblin, H.~S.~Reall, H.~a.~Shinkai and T.~Shiromizu,
  Phys.\ Rev.\  D {\bf 63}, 064015 (2001)
  [arXiv:hep-th/0008177].

\bibitem{Shiromizu:2000pg}
  T.~Shiromizu and M.~Shibata,
  Phys.\ Rev.\  D {\bf 62}, 127502 (2000)
  [arXiv:hep-th/0007203].

\bibitem{Tamaki:2003bq}
  T.~Tamaki, S.~Kanno and J.~Soda,
  Phys.\ Rev.\  D {\bf 69}, 024010 (2004)
  [arXiv:hep-th/0307278].

\bibitem{Kudoh:2003xz}
  H.~Kudoh, T.~Tanaka and T.~Nakamura,
  Phys.\ Rev.\  D {\bf 68}, 024035 (2003)
  [arXiv:gr-qc/0301089].

\bibitem{Csaki:2000dm}
  C.~Csaki, J.~Erlich and C.~Grojean,
  Nucl.\ Phys.\  B {\bf 604}, 312 (2001)
  [arXiv:hep-th/0012143].

\bibitem{Stoica:2001qe}
  H.~Stoica,
  JHEP {\bf 0207}, 060 (2002)
  [arXiv:hep-th/0112020].

\bibitem{Libanov:2005yf}
  M.~V.~Libanov and V.~A.~Rubakov,
  JCAP {\bf 0509}, 005 (2005)
  [arXiv:astro-ph/0504249].

\bibitem{Libanov:2005nv}
  M.~V.~Libanov and V.~A.~Rubakov,
  Phys.\ Rev.\  D {\bf 72}, 123503 (2005)
  [arXiv:hep-ph/0509148].

\bibitem{Bertolami:2006bf}
  O.~Bertolami and C.~Carvalho,
  Phys.\ Rev.\  D {\bf 74}, 084020 (2006)
  [arXiv:gr-qc/0607043].

\bibitem{Ahmadi:2006cr}
  F.~Ahmadi, S.~Jalalzadeh and H.~R.~Sepangi,
  Class.\ Quant.\ Grav.\  {\bf 23}, 4069 (2006)
  [arXiv:gr-qc/0605038];

\bibitem{Ahmadi:2007tr}
  F.~Ahmadi, S.~Jalalzadeh and H.~R.~Sepangi,
  Phys.\ Lett.\  B {\bf 647}, 486 (2007)
  [arXiv:gr-qc/0702103].

\bibitem{Koroteev:2009xd}
  P.~Koroteev and M.~Libanov,
  Phys.\ Rev.\  D {\bf 79}, 045023 (2009)
  [arXiv:0901.4347 [hep-th]];

\bibitem{Farakos:2009ka}
  K.~Farakos, N.~E.~Mavromatos and P.~Pasipoularides,
  J.\ Phys.\ Conf.\ Ser.\  {\bf 189}, 012029 (2009)
  [arXiv:0902.1243 [hep-th]].


\bibitem{Farakos:2009ui}
  K.~Farakos,
  JHEP {\bf 0908}, 031 (2009)
  [arXiv:0903.3356 [hep-th]].


\bibitem{Arianto:2009wc}
  Arianto, F.~P.~Zen and B.~E.~Gunara,
  arXiv:0904.3899 [hep-th].

\bibitem{Kanti:2001vb}
  P.~Kanti, R.~Madden and K.~A.~Olive,
  Phys.\ Rev.\  D {\bf 64}, 044021 (2001)
  [arXiv:hep-th/0104177].

\bibitem{Charmousis:2004zd}
  C.~Charmousis and U.~Ellwanger,
  JHEP {\bf 0402}, 058 (2004)
  [arXiv:hep-th/0402019].

\bibitem{CuadrosMelgar:2005ex}
  B.~Cuadros-Melgar and E.~Papantonopoulos,
  Phys.\ Rev.\  D {\bf 72}, 064008 (2005)
  [arXiv:hep-th/0502169].

\bibitem{Papantonopoulos:2006uj}
  E.~Papantonopoulos,
  arXiv:gr-qc/0601011.

\bibitem{Chatillon:2006vw}
  N.~Chatillon, C.~Macesanu and M.~Trodden,
  Phys.\ Rev.\  D {\bf 74}, 124004 (2006)
  [arXiv:gr-qc/0609093].

\bibitem{Yamauchi:2007wm}
  D.~Yamauchi and M.~Sasaki,
  Prog.\ Theor.\ Phys.\  {\bf 118}, 245 (2007)
  [arXiv:0705.2443 [gr-qc]].

\bibitem{Kanno:2007wj}
  S.~Kanno, D.~Langlois, M.~Sasaki and J.~Soda,
  Prog.\ Theor.\ Phys.\  {\bf 118}, 701 (2007)
  [arXiv:0707.4510 [hep-th]].

\bibitem{Minamitsuji:2008mn}
  M.~Minamitsuji,
  Phys.\ Lett.\  B {\bf 666}, 404 (2008)
  [arXiv:0805.3818 [hep-th]].


\end{thebibliography}
\end{document}